\documentclass{PoS}
\usepackage[english]{babel}
\usepackage[utf8]{inputenc}
\usepackage{cite}
\usepackage{xspace}
\usepackage[T1]{fontenc} % if needed
\usepackage{subfig}
\usepackage{tabularx}  % needed for tables
\usepackage{longtable}  % 
\usepackage{booktabs} % 
\usepackage{units} %
\usepackage{comment}
\usepackage{aas_macros} % needed to resolve ADS journal abbreviations
\usepackage{lineno}
\usepackage{sidecap}
\usepackage{graphicx}
\sidecaptionvpos{figure}{c}

\newcommand{\epem}{$e^+e^-$\xspace}
\newcommand{\gr}{$\gamma$-ray\xspace}

\newcommand{\Grs}{$\gamma$~rays\xspace}
\newcommand{\gammapy}{\textsc{gammapy}\xspace}
\newcommand{\ctools}{\textsc{ctools}\xspace}

\def\LIV{\ifmmode {\mathrm{LIV}}\else{\scshape LIV}\fi\xspace}
\def\LV{\ifmmode {\mathrm{LIV}}\else{\scshape LIV}\fi\xspace}
\def\LI{\ifmmode {\mathrm{LI}}\else{\scshape LI}\fi\xspace}

\title{Testing cosmology and fundamental physics with the Cherenkov Telescope Array}

\ShortTitle{Testing cosmology and fundamental physics with CTA}
        
\author{
    \speaker{H. Martínez-Huerta}$^{\rm a}$,
    J. Biteau$^{\rm b}$, 
    J. Lefaucheur$^{\rm c}$, 
    M. Meyer$^{\rm d}$, 
    S. Pita$^{\rm e}$, 
    and I. Vovk$^{\rm f}$  \ \ \ \ \ \ \ \ \ \ \ \ \ 
    for the CTA Consortium\footnote{For consortium list see PoS(ICRC2019)1177} \\ \\
    $^{\rm a}$ Instituto de Física de São Carlos, Universidade de São Paulo, São Carlos, SP, Brasil \\
    $^{\rm b}$ Institut de Physique Nucléaire, IN2P3/CNRS, Université Paris-Sud, Université Paris-Saclay, Orsay, France\\
    $^{\rm c}$ IRFU, CEA, Université Paris-Saclay, F-91191 Gif-sur-Yvette, France \\
    $^{\rm d}$ Kavli Institute for Particle Astrophysics and Cosmology, Department of Physics and SLAC National Accelerator Laboratory, Stanford University, Menlo Park CA, USA \\ 
    $^{\rm e}$ APC, Univ Paris Diderot, CNRS/IN2P3, CEA/lrfu, Obs de Paris, Sorbonne Paris Cité, France \\
    $^{\rm f}$ Max-Planck-Institut für Physik, München, Germany \\
        E-mail:
        \email{humbertomh@ifsc.usp.br},
        \email{biteau@ipno.in2p3.fr},
        \email{jlefaucheur.pro@gmail.com},
        \email{mameyer@stanford.edu},
        \email{pita@apc.in2p3.fr},
        \email{Ievgen.Vovk@mpp.mpg.de}
        }

\abstract{The Cherenkov Telescope Array (CTA) is the next generation ground-based observatory for \gr astronomy at energies above 30 GeV. Thanks to its unique capabilities, CTA observations will address a plethora of open questions in astrophysics ranging from the origin of cosmic messengers to the exploration of the frontiers of physics. In this note, we present a comprehensive sensitivity study to assess the potential of CTA to measure the \gr absorption on the extragalactic background light (EBL), to constrain or detect intergalactic magnetic fields (IGMFs), and probe physics beyond the standard model such as axion-like particles (ALPs) and Lorentz invariance vio\-lation (LIV), which could modify the \gr spectra features expected from EBL absorption. Our results suggest that CTA will have unprecedented sensitivity to detect IGMF signatures and will probe so-far unexplored regions of the LIV and ALP parameter space. Furthermore, an indirect measurement of the EBL and of its evolution will be performed with unrivaled precision.}

\FullConference{36th International Cosmic Ray Conference -ICRC2019-\\
		July 24th - August 1st, 2019\\
		Madison, WI, U.S.A.}

\begin{document}

%\linenumbers

\section{Introduction}

The Cherenkov Telescope Array (CTA) is the next generation ground-based observatory for \gr astronomy at very-high energies. Exploiting the imaging atmospheric Cherenkov techinque (IACT), it will be capable of detecting \gr in the energy range from 30 GeV to more than 300 TeV with unprecedented precision in energy and directional reconstruction. With more than 100 telescopes, CTA will be located in the northern hemisphere at La Palma, Spain, and in the southern at Paranal, Chile, to provide full sky coverage~\cite{2017arXiv170907997C}. 
The CTA consortium has prepared a Core Program of science observations which makes up ${\sim}\unit[40]{\%}$ of the available observing time in the first ten years of CTA  operations.   This program consists of several Key Science Projects (KSPs), each with its time allocation and different science topics~\cite{2017arXiv170907997C}. The AGN KSP, through observations of Active Galactic Nuclei, will deal with the three key CTA science challenges: probing extreme environments, understanding the origin and role of relativistic cosmic particles, and exploring frontiers in physics. Its strategy involves three different observing programs:  the long-term monitoring program,  the search and follow-up of AGN flares, and a program devoted to high-quality spectra of carefully selected AGNs with systematic coverage of redshift and typology. Consequently, CTA will detect hundreds of blazars at very-high energies, allowing a significant growth of the burgeoning field of \gr cosmology~\cite{Gate:2017plz}. 

In the next sections, we present preliminary studies to assess the potential of CTA to measure the \gr absorption on the extragalactic background light (EBL) in Sec.~\ref{Sec:EBL}, to constrain or detect intergalactic magnetic fields (IGMFs) in Sec.~\ref{Sec:IGMF}, and to probe physics beyond the standard model such as axion-like particles (ALPs) in Sec.~\ref{Sec:ALP} and Lorentz invariance violation (LIV) in Sec.~\ref{Sec:LIV}.

\section{The extragalactic background light}\label{Sec:EBL}

Very-high-energy $\gamma$ rays that propagate from extragalactic sources suffer significant attenuation due to the interaction with the intergalactic background light ($\gamma_{b}$) through the pair production process $\gamma~\gamma_{b} \longrightarrow e^+ e^-$. Due to the energy dependency of this interaction\footnote{In particular, its energy threshold is given by 
$E_{\gamma} \geq 210 \ {\rm GeV}\times\left(\frac{\lambda}{1{\mu m}}\right)$, where $E_{\gamma}$ and $\lambda$ are the energy of the $\gamma$ ray and the wavelength of the low-energy background photon, respectively, in the comoving cosmological frame~\cite{DeAngelis:2013jna}.}, $\gamma$ rays, depending on their energies, are absorbed by the EBL from far UV to far IR. This diffuse light, the second most intense intergalactic background radiation after the CMB, has two components: a first peak from near UV to near IR, which corresponds mainly to the light directly emitted by stars and galaxies since the reionization epoch, and a second peak from near IR to far IR, which corresponds to the fraction of this light which has been absorbed by dust in the interstellar medium and around active galactic nuclei and subsequently reradiated at lower energies. Hence, the EBL carries essential information on the radiation history of the Universe and is a key observable for the modeling of reionization as well as galaxy formation and evolution. In addition, as discussed in the following sections, a good knowledge of EBL is a prerequisite for several topics related to high-energy astrophysics and fundamental physics.

The EBL intensity, its spectral dependence, and its evolution with redshift are today poorly constrained. The estimations derived from galaxy counts in deep surveys are limited, in particular, through the sensitivity of these surveys and should be considered as lower limits, until convergence at the low-end of the luminosity functions is reached at all wavelengths. Direct measurements suffer from contamination by strong foregrounds, in particular, the zodiacal light from the dust in the solar system, and should be considered as upper-limits. A third approach is provided by \gr astronomy~\cite{REF::NIKISHOV::JETP1962} and is based on the imprint on $\gamma$-ray spectra of the $\gamma-\gamma$ absorption, via the attenuation factor $\exp(-\tau(E_{\gamma},z))$, where $\tau$ is the \gr optical depth, $E_{\gamma}$ the \gr energy and $z$ the redshift of the source. This method has been extensively used during the past decade by \gr observatories on the ground (H.E.S.S., MAGIC, VERITAS) or in orbit (\emph{Fermi}-LAT). They converge on compatible measurements of the EBL density in the nearby Universe from UV to near IR with uncertainties better than $20\%$ and provide first constraints on the redshift dependence of the EBL density \cite{biteau2015}.

\begin{SCfigure}[][t]
  \centering
  \includegraphics[width=0.66\textwidth]{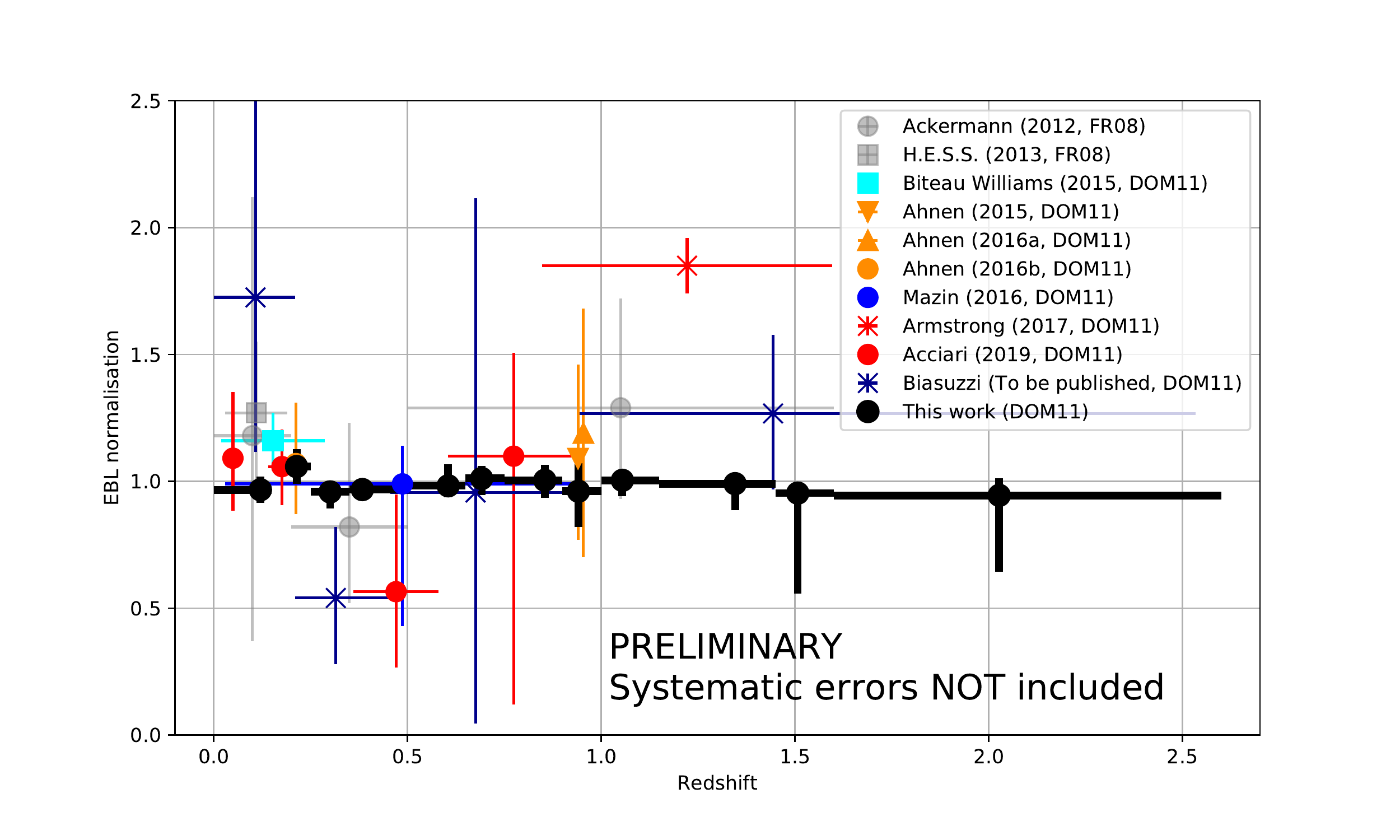}
  \caption{\small Reconstructed EBL scale factor as a function of redshift, considering for each source participating to the joint-fit an exponential cut-off at 1/(1+z) TeV. Constraints currently available in the literature are also shown.} %TODO: add references + merge figures ? }
  \label{Fig:ELB}
\end{SCfigure}

CTA, with its unprecedented sensitivity and extended energy range from $\sim30\,$GeV to a few hundreds of TeV, will have the capacity to measure, for numerous blazars at different redshifts, high-quality spectra including at the same time parts of the spectra not affected or strongly affected by EBL absorption. We estimated the CTA capabilities to reconstruct the intensity of the EBL density, using a simulated dataset of the observation of a hundred of blazars corresponding to ${\sim}\unit[40]{\%}$ of CTA extragalactic Core Program time in the first ten years of CTA operations. These blazars were selected based on their flaring or their long-term behavior as seen by the current generation of IACTs and by \emph{Fermi}-LAT, and their intrinsic spectra were estimated. To be used in the simulations, an additional exponential cut-off at $1/(1+z)$ TeV was considered. The absorption by the EBL was accounted for in the simulations by using the model of Ref.~\cite{Dominguez11}.

The selected sources have a redshift distribution which spreads from $z\sim0$ to $z\sim2.6$.  
We divided them in redshift intervals, considering first a size of $\Delta z=0.05$ for the redshift bins and merging them until a minimal number of 5 sources was reached.
For each redshift interval, we used the collection of signal and background counts obtained in the simulations of the CTA observations, using the latest instrument response functions\footnote{Hereafter, the so-called \texttt{prod3b} instrument response functions are used for both the Northern (La Palma) and Southern (Paranal) CTA sites. $^3$\href{http://cxc.cfa.harvard.edu/contrib/sherpa/}{http://cxc.cfa.harvard.edu/contrib/sherpa/}}.
%including the latest versions of the instrumental response functions (IRFs, ) 
We then fitted the intrinsic spectral parameters of each source and, jointly for all sources, a scaling factor $\alpha$ for the EBL density spectrum of Ref.~\cite{Dominguez11}. The simulations and fits were performed using the Python libraries \gammapy~\cite{Deil2017} and \textsc{sherpa}$^3$.%\footnote{\href{http://cxc.cfa.harvard.edu/contrib/sherpa/}{http://cxc.cfa.harvard.edu/contrib/sherpa/}}.

The results obtained for the reconstruction of the EBL normalization scale $\alpha$ as a function of $z$ are shown in Fig.~\ref{Fig:ELB}. We see that $\alpha$ is very well determined with statistical errors at the level of a few percent for most of the bins. Beyond $z\sim1.4$, the exponential cut-off becomes hard to reconstruct, leading to an enlargement of the error of $\alpha$ towards low values.

In conclusion, CTA should be able to provide a measurement of the EBL density at $z\sim0$ and its evolution with $z$ up to $z\sim2.6$ with unprecedented accuracy. The statistical precision should be better than $10\%$ at least up to $z\sim1$. The systematic uncertainties are currently under estimation.

%==================
\section{The intergalactic magnetic field}\label{Sec:IGMF}

The \gr interaction with the background light results in the production of an \epem pair, which, as charged particles, are sensitive to the intergalactic magnetic field. The cumulative effect of this phenomenon results in an electromagnetic cascade and a secondary \gr flux. 
Measurements may reveal this secondary components through various aspects of the IGMF influence, such as a time delay, the presence of broad spectral features, and an extended halo-like emission around point-like primary sources. 
The strength and coherence length of the IGMF are to date unknown~\cite{2013A&ARv..21...62D}. 
However, current VHE observatories \cite{2014A&A...562A.145H} and simultaneous analysis of IACT and \emph{Fermi}-LAT data \cite{2018ApJS..237...32A} have constrained the IGMF strength to $B_\mathrm{IMGF} \gtrsim 10^{-15}-10^{-14}\,$G, for a coherence length larger than $1\,$Mpc. Given the large collection area and low energy threshold of CTA, its observations promise to constrain or detect these IGMF effects. 

\begin{SCfigure}[][h]
  \centering
  \includegraphics[width=0.54\textwidth]{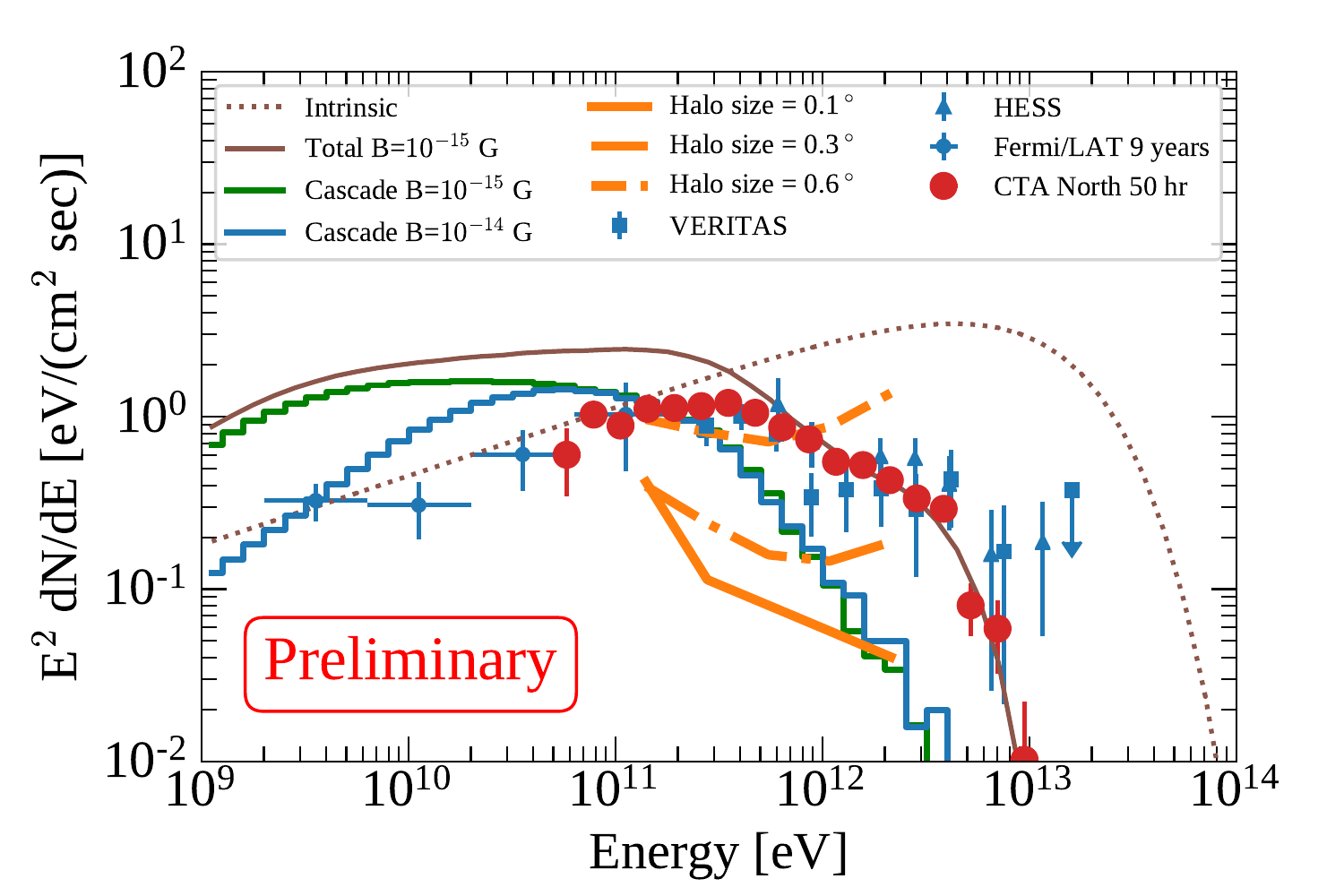}
  \caption{\small
    CTA spectrum of 1ES~0229+200 (in red), simulated for 50~hours of observations with CTA~North. The spectrum is simulated for zero IGMF and thus represents a sum of the intrinsic source (brown dotted line) and the corresponding cascade (not shown) fluxes. Orange solid, dot-dashed, and dashed lines show CTA sensitivity for cascade emission of various extensions.
  }
  \label{Fig:IGMF}
\end{SCfigure}

As an example of the CTA potential, in Fig.~\ref{Fig:IGMF}, we show the simulated 50hrs of CTA observations of the extreme blazar 1ES 0229+200 at $z=0.14$, for which \ctools software is used~\cite{ctools}. We assume that the intrinsic source spectrum follows the power law with exponential cut-off,
%\begin{equation}\label{eq:ple}
  $dN/dE = N_{0} (E/E_0)^{-\Gamma} \exp{(-E/E_\mathrm{cut})}$,
%\end{equation}
where $\Gamma=1.6$ and $E_\mathrm{cut} = 10$~TeV, in agreement with current observations.  The simulation of the electromagnetic cascade is performed with the \textsc{CRPropa} code~\cite{CRPropa} and assuming the EBL model of Ref. \cite{Dominguez11}.  A jet opening angle of $5\deg$ is considered and we assume here that the blazar has been emitting \Grs for more than $10^7$ years. Results for $B_\mathrm{IGMF}\gtrsim 10^{-14}$\,G (blue line) and $10^{-15}$G (gree line) are shown in the Fig.~\ref{Fig:IGMF}.  The excess of this secondary \gr is visible below $\sim 100$ GeV in comparison to the simulated spectrum without the cascade contribution (dotted brown line). Moreover, the cascade suppression is noticeable in the CTA energy range for $B_\mathrm{IGMF}\gtrsim 10^{-14}$\,G.

Besides the presence of broad spectral features due to the cascade contribution, a piece of indubitable evidence for the presence of a strong IGMF would be the detection of degree-scale, extended \gr halos around distant blazars. The presence of existing halos due to an IGMF has already been searched with current IACTs~\cite{HEGRA_IGMF}, nevertheless, because the size of the halo depends on the strength of the magnetic field, the better angular resolution of CTA will enable the search for smaller halos. A characteristic angular spread of the cascade caused by the intervening IGMF is $\theta \simeq 0.5 (B_\mathrm{IGMF}/10^{-14}\mathrm{G})$\,deg at 100\,GeV~\cite{NeronovSemikoz09}. In Fig.~\ref{Fig:IGMF}, we also show examples of CTA 50 hrs sensitivity for cascade emission with halo sizes of  $0.1º, 0.3º$ and $0.6º$  by simulating a point-like 1ES 0229+20.

% --------------------------------

\section{Axion-like particles}\label{Sec:ALP}

Very-high energy \Grs that propagate from distant sources can also probe physics beyond the standard model. In particular, \Grs could oscillate into axion-like particles (ALPs) in the presence of a magnetic field.
ALPs are spin-zero pseudo-Nambu-Goldstone Bosons and result from the breaking of an additional fundamental gauge symmetry in the standard model (SM). ALPs can also arise in extra-dimension scenarios as Kaluza-Klein zero modes of compactified string theories~\cite{2018PrPNP.102...89I}. 
Due to non-perturbative effects or explicit symmetry breaking, ALPs may acquire some mass, $m_a$. Furthermore, in their interactions with SM particles, ALPs may couple to photons through with a coupling strength $g_{a\gamma}$. 
Both parameters are usually taken as independent and they define the ALP parameter space. Interestingly,  ALPs are dark matter candidates if they are sufficiently light and produced non-thermally in the early Universe~\cite{abbott1983}.  

Once \Grs oscillate into ALPs, they evade the pair production process and thus can significantly reduce the effective optical depth, leading to unique features in the spectra of active galaxies.
Evidence for such a reduction has been searched for in blazar observations~\cite{deangelis2007}. 
Previous works have shown that such an effect could be addressed with CTA~\cite{meyer2014cta}. 
However, recent analyses found that \gr spectra are in general compatible with predictions from EBL 
attenuation~\cite{sanchez2013,biteau2015}. The strongest bounds on the photon-ALP coupling for $m_a$ between $\sim4\,\mathrm{neV}$ and $\sim100\,\mathrm{neV}$ are given by the observations of PKS\,2155-304 and the radio galaxy NGC\,1275 with the \emph{Fermi}-LAT and the H.E.S.S. telescopes, respectively \cite{hess2013:alps}. 

\begin{SCfigure}[][t]
\centering
\includegraphics[width = 0.47\textwidth]{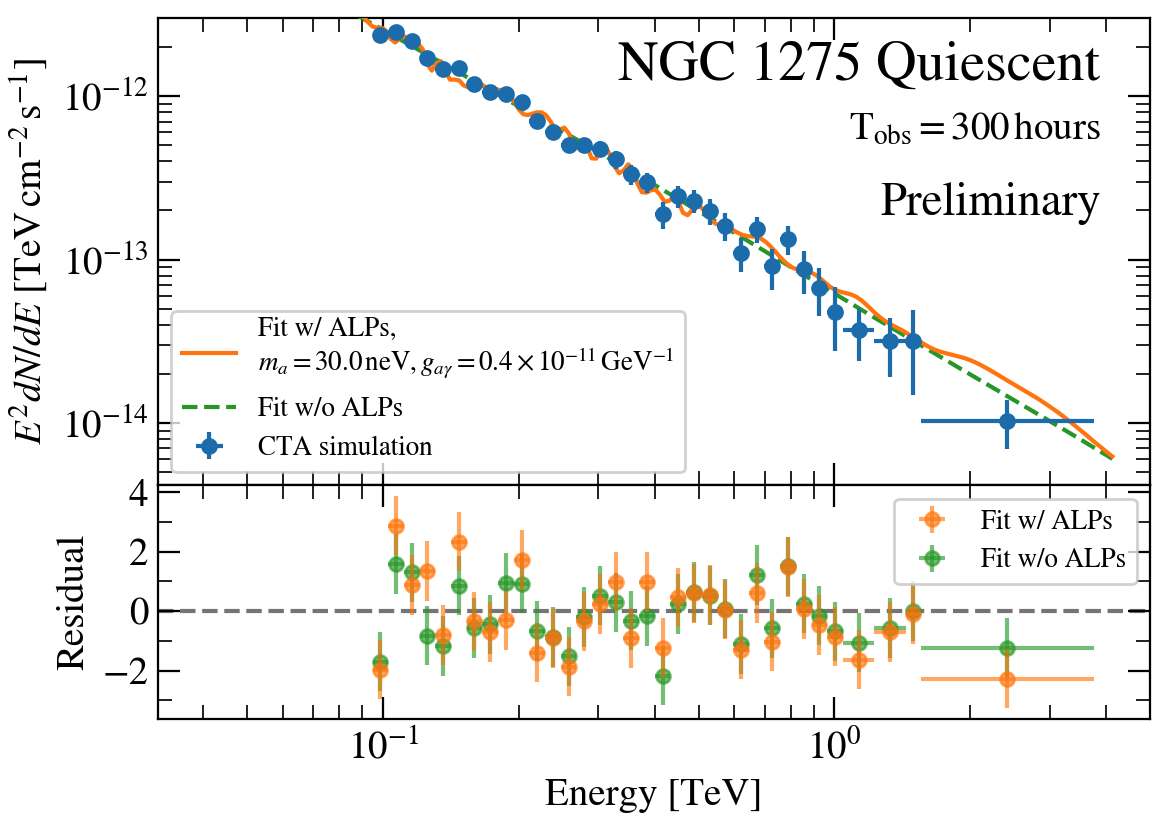}
\caption{\label{Fig:ALP} \small Simulated CTA observations of NGC\,1275 in a quiescent state. The observation is simulated without an ALP effect and fitted both with and without ALPs. 
}
\end{SCfigure}

As an example of the CTA potential to constrain the ALP parameter space, in Fig.~\ref{Fig:ALP} we show a \gammapy simulation for a CTA North observation (zenith angle of $20^\circ$) of NGC\,1275, the central galaxy of the Perseus cluster.
We assume the source in a quiescent state, as measured with the MAGIC telescopes \cite{magic-preseus2016}, with a power-law spectrum with normalization $N_0 = 2.1\times10^{-11}\,\mathrm{TeV}^{-1}\,\mathrm{cm}^{-2}\,\mathrm{s}^{-1}$, at energy $E_0 = 0.2\,\mathrm{TeV}$ and index $\Gamma = 3.6$. The simulation considers EBL absorption with $z \sim 0.018$, an observation time of 300 hours (as planned in the galaxy cluster KSP), and no ALP effect. However, both corresponding spectral fits, with and without ALP effect, are shown in Fig.~\ref{Fig:ALP}. 
The chosen ALP parameters correspond to values where ALPs could constitute almost all dark matter content of the Universe, and a single random realization of the turbulent magnetic field of the Perseus cluster was used. 
By maximizing the logarithm of the Poisson likelihood summed over all energy bins, our preliminary results suggest that CTA will be able to start to probe the ALP dark matter parameter space between $\sim20$\,neV and $\sim100$\,neV especially if CTA observes NGC\,1275 in a flaring state as observed with VERITAS \cite{2017ATel.9931....1M}. 
A previous analysis with a CTA simulation of NGC\,1275, including ALPs effect, studied the CTA potential to probe new regions of the ALP parameter space~\cite{Gate:2017plz}.

%--------------------------

\section{Lorentz invariance violation}\label{Sec:LIV}

High-sensitivity $\gamma$-ray observations at the highest energies can also be used as a test for fundamental physics, such as the Lorentz invariance (\LI). Like any other fundamental principle, exploring its limits of validity has been an essential motivation for theoretical and experimental research~\cite{Colladay:1998}. The formulation of the quantum theory of gravity is one of the main challenges of physics, and some approaches predict a violation of Lorentz invariance (\LIV) at high energy scales~\cite{Alfaro:2005}. 
Although LIV signatures are expected to be small,  some effects of \LIV are expected to increase with energy and distance due to cumulative processes; therefore, astroparticle scenarios provide an unprecedented opportunity for this task due to their very high energies and long distances. Moreover, the CTA unprecedented precision in energy and directional reconstruction will generate an extraordinary opportunity for LIV tests.

A phenomenological generalization of the LIV effects converges on the introduction of a ge\-ne\-ral modification of the relation between energy and momentum. The modified dispersion relation (MDR) can be induced by the introduction of a Lorentz violating term in the Lagrangian or a spontaneous Lorentz symmetry breaking~\cite{Colladay:1998}. The derived physics from such corrections can lead to shifts at the minimum background photon energy that allows the pair production process, given~by
\begin{equation}\label{eq:LIV_th}
\epsilon^{th} = \frac{m_e^2}{4E_{\gamma} K(1-K)} - \frac{1}{4} \delta^{tot}_n \ E_{\gamma}^{n+1},
\end{equation}
where $K$ is the inelasticity of the process, $n$ is the leading order of the LIV correction in the MDR, and $\delta^{\rm tot}_n$ is a linear combination of the \LIV coefficients from the different particle species, $\delta_{a,n}$~\cite{MH_2016azo}. In some effective field theories, $\delta_{a,n}=\zeta_a^{(n)}/M$, where $M$ is the energy scale of the new physics, such as the Plank energy scale, $E_{Pl}\sim$~$10^{28}$~eV, or some Quantum Gravity energy scale, $E_{QG}$, and $\zeta^{(n)}$ are \LIV coefficients. For simplicity, only subluminal photon \LIV is considered ($\delta_{n}^{tot}<0$). In addition, the \LIV correction is taken as $|\delta_{n}| = (E_{\LIV}^{(n)})^{-n}$. The subluminal \LV effect forecast a recovery in the spectrum of TeV-sources that can be measured by the current \gr telescopes~\cite{Stecker:2003,biteau2015}. Although, no \LIV signal of this type has been reported, best 2 sigma limits on $E_{LIV}$ where found to be $E_{\LIV}^{(1)}\ge 12 \times 10^{28}$~eV and $E_{\LIV}^{(2)}\ge 2.4 \times 10^{21}$~eV~\cite{Lang:2018yog}.

\begin{figure}[h]
    \centering
    \subfloat[]{{\includegraphics[width=.47\linewidth]{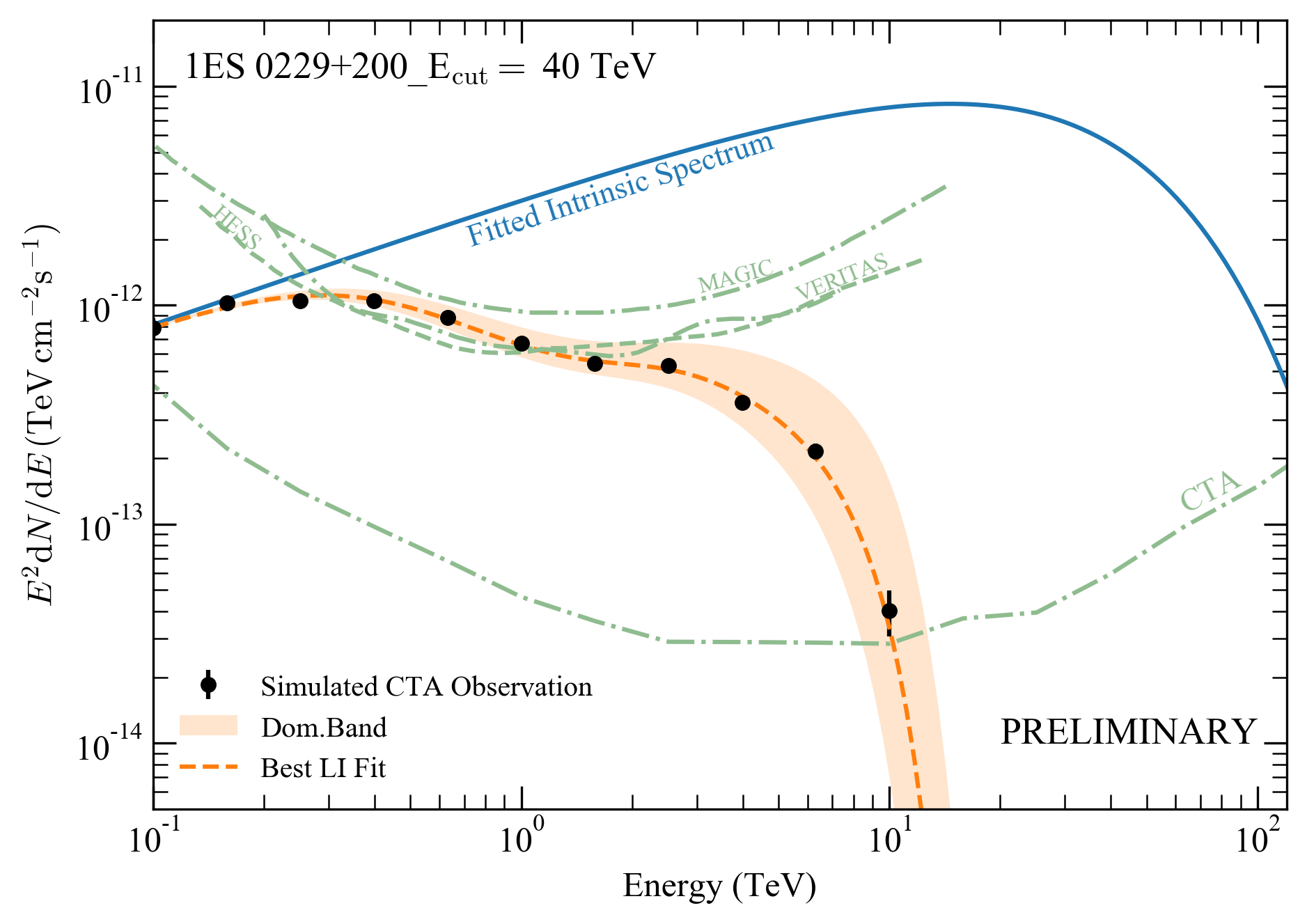} }}
    \subfloat[]{{\includegraphics[width=.47\linewidth]{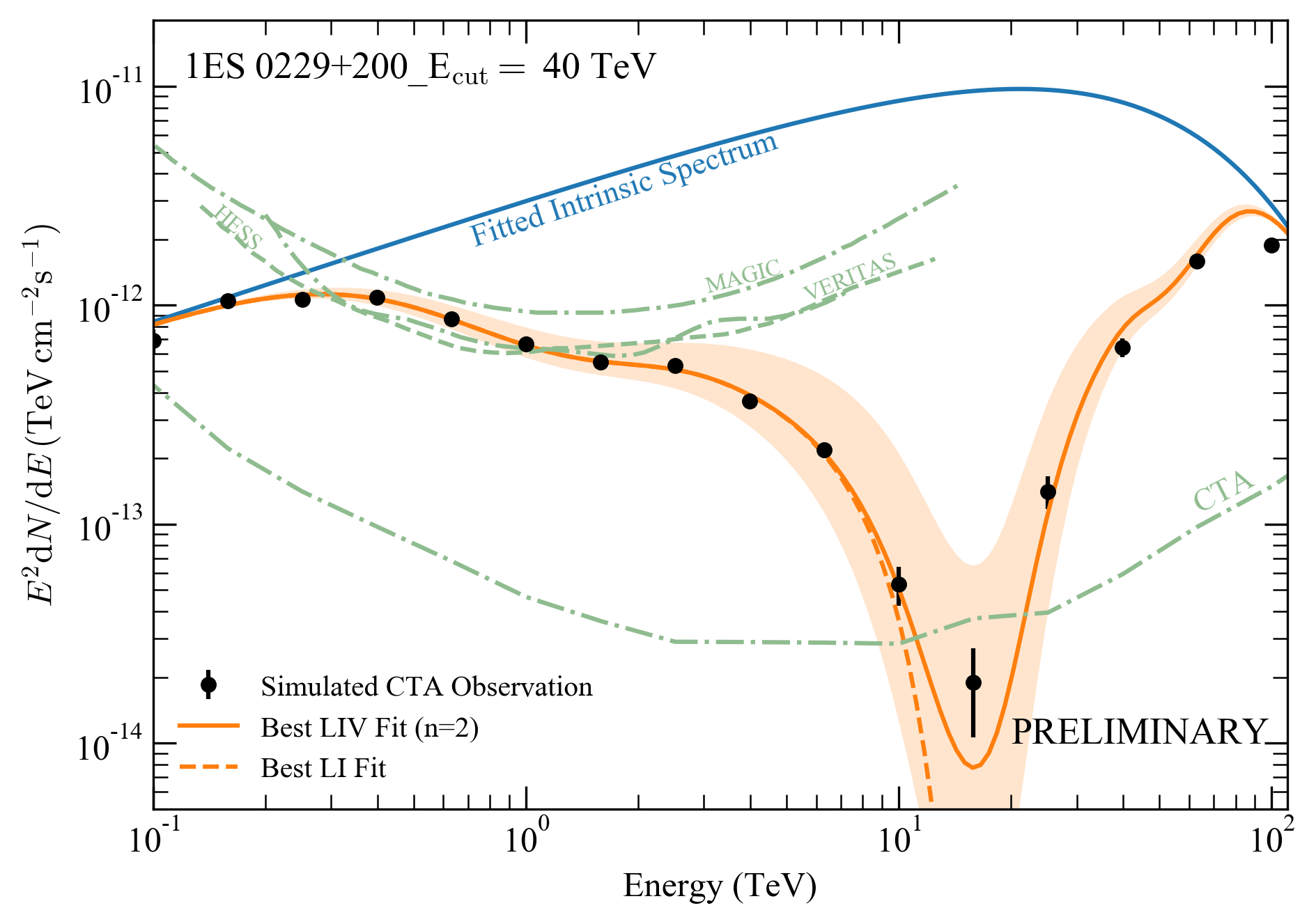} }}%
    \caption{\small %CTA sensitivity to the detection of a LIV signature. Left: 
    The simulated CTA observations of 1ES0229+200 with $E_\mathrm{cut} = 40$ TeV without LIV effect (left) and with the LIV scenario $n = 2$ (right). The shaded band is the variation in the limits of the EBL model. 
    %The orange band is the variation in the upper and lower limits of the  EBL model.
    }
    \label{Fig:LIV}
\end{figure}

As an example of the CTA potential to test \LIV, in Fig.~\ref{Fig:LIV} there are \gammapy simulations for CTA North observations of 1ES 0229+200 as black points, assuming the intrinsic source spectrum in Sec.~\ref{Sec:IGMF}, where $N_{0}~=~1.45\times 10^{-12} {\rm \ TeV^{-1}cm^{-2}s^{-1}}$,  $E_{0}~=~ 1.6$ TeV, $\Gamma~=~1.45$ and $E_{\rm cut}= 40\,$TeV. Once again, the EBL model from Ref.~\cite{Dominguez11} is assumed. 
In Fig.~\ref{Fig:LIV}(a), results without the \LIV effect are shown, while  Fig.~\ref{Fig:LIV}(b) includes the \LIV effect for the scenario where $n=2$ and $\rm E_{\LIV}^{(2)}= 5\times 10^{20}$ eV, through Eq.~(\ref{eq:LIV_th}). We found similar effects for $\rm E_{\LIV}^{(1)}\sim \rm E_{Pl}$.  For comparison, the current IACT (HESS, MAGIC, and VERITAS) and CTA sensitivity thresholds are shown in the dashed-dotted lines. It is clear from Fig.~\ref{Fig:LIV}, that CTA will improve the possibility to detect a LIV signal from the previous IACT. In the particular scenario with the \LIV effect in Fig.~\ref{Fig:LIV}(b), the recovery in the spectrum at TeV energies would be clearly detected by CTA. Preliminary results from the simulations of CTA observations of the nearby blazar Mrk 501 can be found in Ref. \cite{Gate:2017plz}.

\section{Conclusions}

We have presented a preliminary study of the CTA sensitivity to signatures imprinted on \gr spectra due to a variety of effects that might affect the propagation of \gr over cosmological distances.  These preliminary results suggest that CTA will be able to provide an indirect measurement of the EBL  and its evolution with unparalleled precision. Besides, CTA will have unprecedented sensitivity to detect IGMF signatures and will probe so-far unexplored regions of the LIV and ALP parameter space with unrivaled precision. A complete study that will address the full potential of CTA in this science topics, including systematic uncertainties and the development of dedicated analysis techniques, is currently in preparation within the CTA Consortium.

\footnotesize{
\section*{Acknowledgements}
This work was conducted in the context of the CTA Physics Working Group. We gratefully acknowledge financial support from the agencies and organizations listed here:  \href{http://www.cta-observatory.org/consortium_acknowledgments}{http://www.cta-observatory.org/consortium$\_$acknowledgments/}. HMH acknowledges FAPESP support No. 2015/15897-1 and 2017/03680-3 and the National Laboratory for Scientific Computing (LNCC/MCTI, Brazil) for providing HPC resources of the SDumont supercomputer (\href{https://sdumont.lncc.br}{sdumont.lncc.br}).}

%\bibliographystyle{unsrt}
%\bibliography{bibfile.bib}

\providecommand{\newblock}{}

\end{document}